# Higgs and Top Quark Masses in the Standard Model without Elementary Higgs Boson

V. N. Gribov
*Institut für Theoretische Kernphysik,*
*Universität Bonn, Nußallee 14-16, 53115 Bonn, Germany*
*and*
*Research Institute for Particle and Nuclear Physics,*
*Budapest, Hungary*
*and*
*L. D. Landau Institute for Theoretical Physics,*
*Moscow, Russia*

## Abstract

In this short note I present a simple calculation of the top quark and Higgs masses, based on the idea that in the standard model without elementary Higgs the fact that the $U(1)_Y$ coupling becomes of the order of unity at the Landau scale $\lambda$ leads to spontaneous symmetry breaking and generation of masses.

## I. INTRODUCTION

The basis of this paper is the idea that in the standard model without elementary Higgs particle the $U(1)_Y$ coupling becomes of the order of unity at the Landau scale $\lambda$, thus resulting in a spontaneous symmetry breaking and generation of masses. In this case the longitudinal components of the $W$, $Z^0$ and the Higgs have to be bound states similar to the deuteron in the zero–radius limit of nuclear forces. The usual $SU(2)$ and $SU(3)$ interactions would act like the electromagnetic interaction between nucleons in the deuteron, and the situation could be treated perturbatively.
The existence of the longitudinal components of the $W$, $Z^0$ is connected to the fact that the generation of quark and lepton masses leads to non–conservation of the two components of the $SU(2)$ left–handed current and the left–handed component of the $U(1)$ current. The existence of the Higgs is due to the degeneracy of states with positive and negative parity at short distances. In a more detailed paper we will show that there is a good chance for such a theory to exist.

## II. THE *W* AND $Z^0$ MASSES

We start with the calculation of the $W$ mass. The fermionic part of the polarization operator of the $SU(2)$ bosons is given by

$$\Pi^{ab}_{\mu\nu} = -\frac{g_2^2}{4}\sum_j \int \frac{d^4q}{(2\pi)^4 i}\,\mathrm{tr}(\tau^a\gamma^L_\mu G_j(q)\gamma^L_\nu\tau^b G_j(q-p)) \tag{1}$$

$=$ 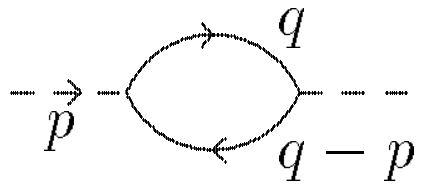

,

where $\gamma^L_\mu = \gamma_\mu \cdot \frac{1}{2}(1-\gamma_5)$ and the sum extends over all fermionic generations. In the absence of quark and lepton masses

$$\Pi^{ab}_{\mu\nu} = \mathring{\Pi}^{ab}_{\mu\nu} = (g_{\mu\nu}p^2 - p_\mu p_\nu)\,\Pi(p^2)\,\delta_{ab} \tag{2}$$

due to current conservation. For massive fermions

$$G_j = \frac{1}{m_j - \not{q}}\,. \tag{3}$$

Here $m_j$ is a function of $\tau_3$ and can be written as

$$m_j = m_{\uparrow j}\cdot\frac{1}{2}(1+\tau_3) + m_{\downarrow j}\cdot\frac{1}{2}(1-\tau_3)\,. \tag{4}$$

Because of (3), the form (2) is no longer valid. Instead, in second order in $m$, $\Pi_{\mu\nu}$ can be written for $q \gg m$ and $p^2 \to 0$ as

$$\Pi^{ab}_{\mu\nu} = \mathring{\Pi}^{ab}_{\mu\nu} - \frac{g_2^2}{2}\sum_j \int_{m_j \ll q \ll \lambda} \frac{d^4q}{(2\pi)^4 i}\,\mathrm{tr}(\gamma^L_\mu \not{q}\, m_j^2 \gamma^L_\nu \not{q})\frac{1}{q^6}\,\delta_{ab} \tag{5}$$

$$= \mathring{\Pi}^{ab}_{\mu\nu} + \Delta\Pi^{ab}_{\mu\nu}\,. \tag{6}$$

Applying (4),

$$\Pi^{ab}_{\mu\nu} = \mathring{\Pi}^{ab}_{\mu\nu} + g_{\mu\nu}\delta_{ab}\frac{3g_2^2}{2\cdot 16\pi^2}\sum_{q,l}\int_{m^2}\frac{dq^2}{q^2}(m^2_{\uparrow q} + m^2_{\downarrow q} + \frac{1}{3}m^2_l)\,. \tag{7}$$

The lower limit in (7) is defined by the physical masses of quarks and leptons. At first sight this integral looks divergent. But if we take into account the dependence of $m$ on $q^2$ the integral turns out to be convergent for the main contribution, given by the top quark mass, if the upper limit is the Landau scale $\lambda$. The dependence of the masses on $q^2$ is well–known in the standard model (see for example [1]). It is defined by the $SU(3)$ and $U(1)$ interactions only and reads (for three generations)

$$m_{\uparrow q}(q^2) = m_{\uparrow q}\left[\frac{\alpha_s(q^2)}{\alpha_s(m^2_{\uparrow q})}\right]^{\frac{4}{7}}\left[\frac{\alpha'(m^2_{\uparrow q})}{\alpha'(q^2)}\right]^{\frac{1}{20}} \tag{8}$$

$$m_{\downarrow q}(q^2) = m_{\downarrow q}\left[\frac{\alpha_s(q^2)}{\alpha_s(m^2_{\downarrow q})}\right]^{\frac{4}{7}}\left[\frac{\alpha'(m^2_{\downarrow q})}{\alpha'(q^2)}\right]^{-\frac{1}{40}} \tag{9}$$

$$m_l(q^2) = m_l\left[\frac{\alpha'(m^2_l)}{\alpha'(q^2)}\right]^{\frac{9}{40}}\,. \tag{10}$$

The dependence on the $U(1)$ coupling $\alpha'$ is very slow and becomes essential only near the Landau scale. We can therefore neglect it when integrating up to the Landau scale. Taking into account only the contribution of the heaviest quark – the top quark – and using the $\alpha_s$ dependence on $q^2$,

$$\alpha_s^{-1}(q^2) = \alpha_s^{-1}(m^2) + \frac{7}{4\pi}\ln\frac{q^2}{m^2}\,, \tag{11}$$

and (8) we obtain

$$\Pi^{ab}_{\mu\nu} = \mathring{\Pi}^{ab}_{\mu\nu} + g_2^2\, g_{\mu\nu}\delta_{ab}\frac{3}{2g_s^2}m_t^2\left\{1 - \left[\frac{\alpha_s(\lambda^2)}{\alpha_s(m_t^2)}\right]^{\frac{1}{7}}\right\}\,, \tag{12}$$

where $g_s^2 = 4\pi\alpha_s(m_t^2)$. After replacing $\delta_{ab}$ by $(\delta_{ab} - \delta_{a3}\delta_{b3})$ we find

$$m_W^2 = g_2^2\frac{v^2}{4} \tag{13}$$

$$\frac{v^2}{4} = \frac{3}{2g_s^2}m_t^2\left\{1 - \left[\frac{\alpha_s(\lambda^2)}{\alpha_s(m_t^2)}\right]^{\frac{1}{7}}\right\}\,. \tag{14}$$

In order to calculate the $Z^0$ mass, we have to consider not only the transition between the zero–components of $W$, given by $\Delta\Pi_{\mu\nu}(2,2)$. We also have to consider the transitions between $W_0$ and the $U(1)$ boson – $\Delta\Pi_{\mu\nu}(1,2)$ and $\Delta\Pi_{\mu\nu}(2,1)$ – and between $U(1)$ bosons, $\Delta\Pi_{\mu\nu}(1,1)$. We can do this very easily when noticing that the $U(1)$ current has the structure

$$\gamma^L_\mu\frac{Y_L}{2} + \gamma^R_\mu\frac{Y_R}{2} = Q\gamma_\mu - \gamma^L_\mu\frac{\tau_3}{2}\,, \tag{15}$$

where $Q$ is the operator of electric charge and $Q\gamma_\mu$ is the conserved electric current. Therefore, because the non–conserved part $\Delta\Pi_{\mu\nu}(2,2)$ is equal

$$\Delta\Pi_{\mu\nu}(2,2) = g_2^2\frac{v^2}{4} \tag{16}$$

then

$$\Delta\Pi_{\mu\nu}(1,2) = \Delta\Pi_{\mu\nu}(2,1) = -g_2\, g'\frac{v^2}{4} \tag{17}$$

$$\Delta\Pi_{\mu\nu}(1,1) = g'^2\frac{v^2}{4} \tag{18}$$

and this will reproduce the usual features of $Z^0$ and $\gamma$ with $v/2 = 123.11$ GeV.

Up to now we did not discuss Goldstone particles. As a result our $\Delta\Pi_{\mu\nu}$ is not transverse. But if mass generation is due to spontaneous symmetry breaking, currents have to be conserved via the existence of Goldstone particles and so $\Delta\Pi_{\mu\nu}$ has to be transverse. In order to find the correct expression for $\Delta\Pi_{\mu\nu}$, we have to add to the left–handed $SU(2)$ current the Goldstone contribution:

$\Gamma^a_\mu =$ 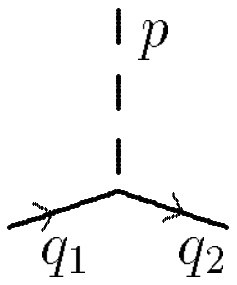

$+$ 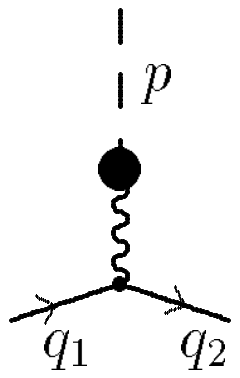

(19)

$$= \gamma^L_\mu\,\frac{\tau^a}{2} - f_{ab}\,g_b\,\frac{p_\mu}{p^2}\,. \tag{20}$$

Here $g_b$ denotes the Goldstone–fermion coupling and $f_{ab}$ is the current–Goldstone transition amplitude. These couplings are defined by the Ward identity for the new current

$$p^\mu\Gamma^a_\mu = \frac{1}{2}(1+\gamma_5)\,\frac{\tau^a}{2}\,G^{-1}(q_1) - G^{-1}(q_2)\cdot\frac{1}{2}(1-\gamma_5)\frac{\tau^a}{2}\,. \tag{21}$$

For $G$ having the form (3) the Ward identity for $p = 0$ gives us

$$\frac{1}{4}[\tau^a, m] + \frac{1}{4}\gamma_5\{\tau^a, m\} = -f_{ab}\,g_b\,. \tag{22}$$

For $f_{ab}$ being a transition amplitude and because we suppose the Goldstone to be a fermionic bound state

$f_{ab}p_\mu = \gamma^L_\mu\frac{\tau^a}{2}$ 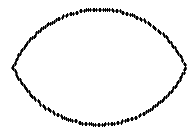 $g^b$ (23)

$$= -\sum_j\int\frac{d^4q}{(2\pi)^4 i}\,\mathrm{tr}(\gamma^L_\mu\frac{\tau^a}{2}\,G_j g^b G_j)\,. \tag{24}$$

Combining (22) and (24) and calculating the integral (24) in the same way as we did in the calculation of $\Delta\Pi_{\mu\nu}$, one can derive

$$if_{ab} = f\delta_{a3}\delta_{b3} + f^{\perp}_{ab} \tag{25}$$

$$f^{\perp}_{ac}f^{\perp}_{cb} = f^2(\delta_{ab} - \delta_{a3}\delta_{b3}) \tag{26}$$

with $f^2 = v^2/4$, as defined in (14) from (7) and (12). Further we can simplify $g_a$ by choosing an appropriate basis in the $ab$–12–plane,

$$g_a = i\left([\tau^a, m] + \gamma_5\{\tau^a, m\}\right)\frac{1}{2f}\,, \tag{27}$$

$$g_3 = i\gamma_5\,\tau_3\,\frac{m}{2f}\,. \tag{28}$$

Now it is easy to write down the correct expression for $\Delta\Pi_{\mu\nu}$. We have to replace both vertices in the diagram of formula (1) by the two terms in (19) or (20). As a result $\Delta\Pi_{\mu\nu}$ would be represented by four diagrams:

$\Delta\Pi_{\mu\nu} =$ 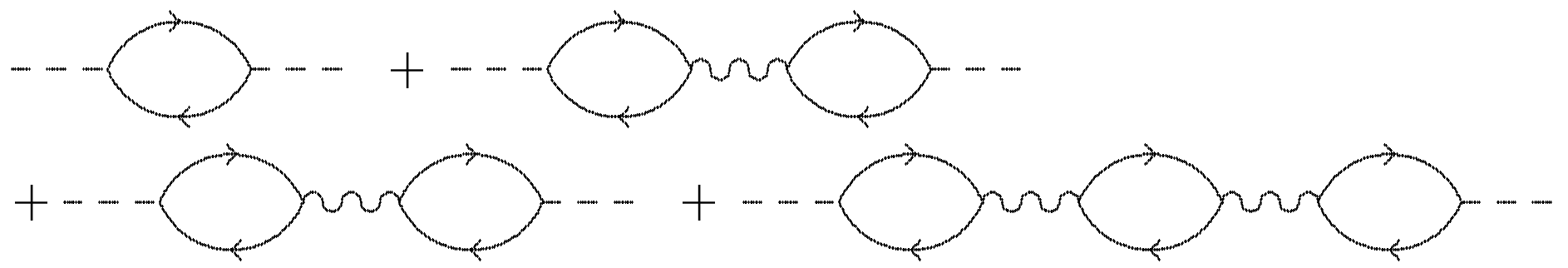 , (29)

where in turn we have replaced 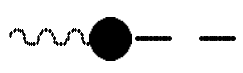 by 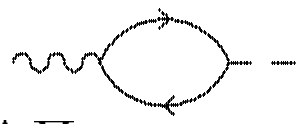 which actually defines it. The first two terms give the correct expression for $\Delta\Pi_{\mu\nu}$

$$\Delta\Pi_{\mu\nu} = f^2\left(g_{\mu\nu} - \frac{p_\mu p_\nu}{p^2}\right)\,. \tag{30}$$

The last two diagrams have to cancel each other. The cancellation condition requires that the new loop in the fourth term of (29) fulfills

$\Sigma^G_{ab} = g_a$ 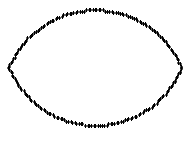 $g_b = p^2\delta_{ab}$ , (31)

which is a natural condition for the self–energy $\Sigma^G$ of a Goldstone type bound state. It is this cancellation condition that will enable us to calculate the Higgs mass.

## III. THE HIGGS MASS

The existence of a Higgs particle is suggested by the fact that at short distances, where all particle masses are zero, bound states have to be degenerate in parity. Therefore, if a pseudoscalar Goldstone particle exists, according to (28), there also has to be a scalar bound state which we will call Higgs particle. Due to large distance effects the mass of this state would not be zero. Because of the degeneracy at short distances, the self–energies of these two states would have the same structure:

$\Sigma_S = g$ 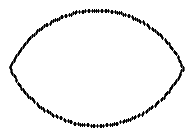 $g$ (32)

$\Sigma_{PS} = i\gamma_5 g\tau_3$ 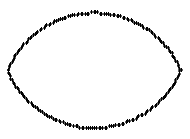 $i\gamma_5 g\tau_3$ (33)

with

$$g_{\rm Higgs} = g \equiv \frac{m}{2f}\,. \tag{34}$$

The mass of the scalar state is defined by the condition

$$\Sigma_S(p^2 = m_H^2) = 0\,. \tag{35}$$

This condition can be written in another form, if we know that the pseudoscalar Goldstone state satisfies (31):

$$\Sigma_S(p^2) - \Sigma_{PS}(p^2) + p^2 = 0 \tag{36}$$

$$\Sigma_S(p^2) = -\sum_j \int \frac{d^4q}{(2\pi)^4 i}\,{\rm tr}(g\,G_j\,g\,G_j) \tag{37}$$

$$\Sigma_{PS}(p^2) = -\sum_j \int \frac{d^4q}{(2\pi)^4 i}\,{\rm tr}(i\gamma_5 g\,G_j i\gamma_5 g\,G_j) \tag{38}$$

so that

$$\begin{aligned} p^2 &= \Sigma_{PS}(p^2) - \Sigma_S(p^2) \\ &= \sum_j \int \frac{d^4q}{(2\pi)^4 i}\,{\rm tr}(g\{\gamma_5, G_j\}\gamma_5 g\,G_j)\,. \end{aligned} \tag{39}$$

The integral on the right–hand side is convergent in the same sense as we discussed before and we find

$$m_H^2 = \frac{2\cdot 3}{f^2\cdot 16\pi^2}\sum_{q,l}\int \frac{dq^2}{q^2}(m_{\uparrow q}^4 + m_{\downarrow q}^4 + \frac{1}{3}m_l^4)\,. \tag{40}$$

Using the expression for $f^2$ (see (14), (7) and (12)) we can write (40) in the form

$$m_H^2 \;=\; 4\,\frac{\sum_{q,l}\int\frac{dq^2}{q^2}(m_{\uparrow q}^4 + m_{\downarrow q}^4 + \frac{1}{3}m_l^4)}{\sum_{q,l}\int\frac{dq^2}{q^2}(m_{\uparrow q}^2 + m_{\downarrow q}^2 + \frac{1}{3}m_l^2)}\,. \tag{41}$$

Neglecting the contributions of all quarks and leptons except the top quark we have

$$m_H^2 \;=\; 4\,\frac{\int_{m_t^2}^{\lambda^2}\frac{dq^2}{q^2}m_t^4(q^2)}{\int_{m_t^2}^{\lambda^2}\frac{dq^2}{q^2}m_t^2(q^2)} \tag{42}$$

or

$$m_H^2 = \frac{3}{2v^2\pi^2}\int_{m_t^2}^{\lambda^2}\frac{dq^2}{q^2}m_t^4(q^2)\,. \tag{43}$$

Using (8) and (11) and neglecting the $\alpha'$ dependence we obtain

$$m_H^2 = \frac{8m_t^4}{3g_s^2 v^2}\left\{1 - \left[\frac{\alpha_s(\lambda^2)}{\alpha_s(m_t^2)}\right]^{\frac{9}{7}}\right\}. \tag{44}$$

Formulas (14) and (44) are the predictions we mentioned in the beginning of the paper. The quantity $\alpha_s(\lambda^2)$ can be expressed through the electromagnetic coupling $\alpha_e(m_t^2)$. From

$$(\alpha'(\lambda^2))^{-1} = (\alpha'(m_t^2))^{-1} - \frac{5}{3\pi}\ln\frac{\lambda^2}{m_t^2}, \tag{45}$$

in order to have $\alpha'(\lambda^2) \sim 1$

$$\ln\frac{\lambda^2}{m_t^2} \approx \frac{3\pi}{5}\frac{1}{\alpha'(m_t^2)}. \tag{46}$$

Using (11) and the relation $\alpha_e(m_t^2) = \alpha'(m_t^2)\cos^2\theta_W$ ($\theta_W$ – Weinberg angle), we have

$$\frac{\alpha_s(m_t^2)}{\alpha_s(\lambda^2)} = 1 + \frac{21}{20}\cos^2\theta_W\,\frac{\alpha_s(m_t^2)}{\alpha_e(m_t^2)}. \tag{47}$$

The concrete values for $m_t$ and $m_H$ depend essentially on the value of the strong coupling at the top quark scale, which is not very well known. For example, for $\alpha_s(m_t^2) = 0.11$

$$m_t = 215 \text{ GeV} \tag{48}$$
$$m_H = 255 \text{ GeV} \tag{49}$$

in contradiction to Fermilab data, and for $\alpha_s(m_t^2) = 0.09$

$$m_t = 201 \text{ GeV} \tag{50}$$
$$m_H = 252 \text{ GeV}. \tag{51}$$

Accepting the Fermilab result $m_t = 174$ GeV [2], we find $m_H = 167$ GeV from (44) for $\alpha_s(m_t^2) = 0.11$.

However, the fact that the masses calculated for the Higgs and top quark very roughly, but essentially from first principles, turn out to be in the correct region seems to me very encouraging. At the same time, there may be an indication that some finite contribution to $v^2/4$ from the region where $q^2$ is of order of $\lambda^2$ could survive. In this respect the prediction of the Higgs mass becomes very important, because it is less dependent on the large momentum region of the theory than the top quark mass.

## IV. CONCLUSION

The idea of a non–elementary nature of the Higgs particle was widely discussed in connection with the so–called top quark condensate [3]. The essential difference to these approaches is that in my discussion the top quark has no special interaction. It becomes important only because it is the heaviest fermion. My Goldstone and Higgs particles consist of all existing quarks and leptons on equal footing. The ratio of the contributions of the different fermions is defined by their masses and the $U(1)$ interaction at short distances.

It is interesting to notice that the Golstone and the Higgs interact weakly with fermions at the scale $m_t$. The largest of the Goldstone couplings – the one to the top – is, according to (14) and (34),

$$\frac{g_t^2}{4\pi} = \frac{m_t^2}{16\pi f^2} = \frac{1}{6} \frac{\alpha_s(m_t^2)}{1 - \left[\frac{\alpha_s(\lambda^2)}{\alpha_s(m_t^2)}\right]^{\frac{1}{7}}} \,. \tag{52}$$

If we accept this formula for an arbitrary scale $m_t \to \mu$, we would find that $g_t^2$ decreases with the momentum scale, similar to the QCD coupling $\alpha_s$. After passing the "grand unification" point $\alpha_s \sim \alpha'$, the behavior of $g_t^2$ changes. It starts to increase with momentum and asymptotically becomes proportional to the $U(1)$ coupling

$$\frac{g_t^2(\mu^2)}{4\pi} = \frac{10}{9}\,\alpha'(\mu^2)\,. \tag{53}$$

The Higgs self–interaction has the usual properties of the Glashow–Weinberg–Salam model, but here there is no question of fine–tuning the Higgs mass, because it is fixed by the symmetry condition (36).

These and other details of the present theory, such as the influence of the Goldstone couplings on the quark mass dependence, will be discussed in the next paper [4].

**Acknowledgments:** I would like to express my deep gratitude to Yu. Dokshitzer for numerous discussions of various aspects of the theory without an elementary Higgs particle. I am also grateful to C. Ewerz for discussions and his help in preparing the manuscript. I wish to thank all the members of the Institute for Theoretical Nuclear Physics for the kind and inspiring atmosphere I experienced during my stay in Bonn.